\documentclass[lettersize,journal]{IEEEtran}
\usepackage{amsmath,amsfonts}
\usepackage{algorithmic}
\usepackage{algorithm}
\usepackage{array}
\usepackage[caption=false,font=normalsize,labelfont=sf,textfont=sf]{subfig}
\usepackage{textcomp}
\usepackage{stfloats}
\usepackage{url}
\usepackage{verbatim}
\usepackage{graphicx}
\usepackage{cite}

\usepackage{comment}
\usepackage{multirow}
\usepackage{array}
\usepackage{booktabs}
\usepackage{amssymb}
\usepackage{pifont}
\usepackage{makecell} 
\usepackage{enumitem}
\usepackage{xcolor}
\usepackage{tabularx}
\usepackage{hyperref}
\usepackage[table]{xcolor}

\newcommand{\cmark}{\ding{51}}
\newcommand{\xmark}{\ding{55}}
\newcommand{\lmark}{\ding{115}}

\definecolor{MyGreen}{HTML}{1B9E77}
\definecolor{MyOrange}{HTML}{E69F00}
\definecolor{MyRed}{HTML}{D55E00}

\definecolor{myPurple}{RGB}{143,0,255}

\newcommand{\gmark}{\textcolor{MyGreen}{\cmark}}
\newcommand{\ymark}{\textcolor{MyOrange}{\lmark}}
\newcommand{\rmark}{\textcolor{MyRed}{\xmark}}

\begin{document}

\title{VoiceDesigner: Text-to-Voice Generation and Editing \\ via Unified Diffusion Modeling and Data Augmentation}

\author{
Jiarui Hai,
Karan Thakkar,
Ke Chen,
Yunyun Wang,
Jiaqi Su,
Rithesh Kumar,
\\
Mounya Elhilali,~\IEEEmembership{Senior~Member,~IEEE},
and Zeyu Jin,~\IEEEmembership{Senior~Member,~IEEE}
\thanks{Work done during Jiarui Hai's internship at Adobe.}
\thanks{Jiarui Hai, Karan Thakkar, and Mounya Elhilali are with Johns Hopkins University. Ke Chen, Yunyun Wang, Jiaqi Su, Rithesh Kumar, and Zeyu Jin are with Adobe Research.}

}

\maketitle

\begin{abstract}
Recent breakthroughs in generative models have made text-to-voice generation (TTV) possible, enabling the synthesis of speech directly from textual voice descriptions. However, existing systems face two key challenges. First, they struggle to generate a diverse range of voices, spanning real-world human speakers and fictional characters. Second, they lack robust and flexible voice editing capabilities, such as voice cloning and the ability to modify attributes like emotion and tone.
In this paper, we propose VoiceDesigner, a unified framework for text-to-voice generation and editing that supports diverse and controllable voice design. To tackle the above challenges, we propose solutions from two perspectives. First, we develop a hybrid data pipeline that leverages digital signal processing techniques and speech generation models to construct a diverse voice dataset covering both real-world and fictional voices. Second, we introduce a diffusion transformer with architectural improvements to better handle complex conditioning and enhance multi-task performance, enabling unified voice generation and editing.
Through subjective and objective evaluations, VoiceDesigner achieves superior prompt alignment with both voice descriptions and editing instructions, while maintaining competitive perceptual quality and voice usability compared to state-of-the-art TTV models.
\end{abstract}


\section{Introduction}
\label{section:introduction}
Text-to-voice generation (TTV) aims to synthesize speech from both the transcript and a textual description of the speaker voice. It extends traditional text-to-speech task (TTS), where the voice is typically fixed to predefined speaker identities, or adapted through audio references for cloning~\cite{chen2025f5, zhou2025indextts2, eskimez2024e2, yang2025simplespeech}. Recent advances in deep learning have enabled the TTV systems, where voice attributes decsribed in natural language can directly guide the speech generation process~\cite{shimizu2024prompttts++, yang2024instructtts, wang2025capspeech}. This emerging task opens up potential applications of voice design in daily content creation, gaming design, and film production. 

Despite this progress, existing TTV models remain limited in the scope of voices they can generate. Most models primarily synthesize realistic human voices characterized by speaker traits (e.g., age, gender, pitch) and style traits (e.g., emotion, projection)~\cite{wang2025capspeech}. While they can effectively handle prompts describing common combinations of these traits, such as \textit{``a young man speaking quickly with excitement''}, they still struggle to reliably generate less common speaking styles, such as nervous or whispering speech.
More importantly, many creative applications, including film and game production, require voices for fictional human and non-human characters (e.g., pirates, rebels, queens, monsters, and demons). Such voices are often specified through character identities and subjective impressions rather than explicit voice traits, as in \textit{``a massive male dragon speaking in a deep, thunderous, rumbling voice''}. These descriptions are more intuitive for users yet difficult to represent using trait-based prompts alone. Existing models perform poorly in these scenarios largely because their training data predominantly consists of natural human speech from sources such as audiobooks and podcasts~\cite{he2024emilia, zen2019libritts}, with limited coverage of diverse fictional or character voices.

Beyond generating diverse voices, practical voice design also requires the ability to reuse and refine a designed voice for different creative needs. However, existing voice cloning systems~\cite{yang2025simplespeech, chen2025f5, zhou2025indextts2, du2025cosyvoice, wu2024laugh} are primarily designed for conventional human voices and are less robust when cloning unconventional or heavily stylized voices. Likewise, instruction-based voice editing methods~\cite{chen2025isse, yan2025step, zhou2025indextts2} exhibit limited editing capability and often struggle to preserve voice quality. Furthermore, voice generation and editing are typically implemented as separate systems, increasing training and deployment costs while limiting their applicability in low-resource settings.

\begin{table}[t]
\centering
\caption{TTV capabilities across models 
(\ymark: limited support).}
\label{tab:ttv_model_comp}

\resizebox{\linewidth}{!}{
\begin{tabular}{lccc|cc}
\toprule

\multirow{3}{*}{\textbf{Model}}
& \multicolumn{3}{c|}{\textbf{Voice Generation Scope}}
& \multicolumn{2}{c}{\textbf{Editing Capability}} \\ \cmidrule{2-6} 

& \makecell{\textbf{Speaker}\\\textbf{Traits}}
& \makecell{\textbf{Style}\\\textbf{Traits}}
& \makecell{\textbf{Character}\\\textbf{Design}}
& \makecell{\textbf{Voice}\\\textbf{Cloning}}
& \makecell{\textbf{Voice}\\\textbf{Editing}} \\

\midrule

PromptTTS \cite{guo2023prompttts, shimizu2024prompttts++}   
& \gmark & \ymark & \rmark & \rmark & \rmark \\

ParlerTTS \cite{lacombe-etal-2024-parler-tts}  
& \gmark & \ymark & \rmark & \rmark & \rmark \\

VoxInstruct \cite{zhou2024voxinstruct} 
& \gmark & \gmark & \rmark & \rmark & \rmark \\

CapSpeech \cite{wang2025capspeech}  
& \gmark & \gmark & \rmark & \rmark & \rmark \\

Qwen3TTS \cite{hu2026qwen3}   
& \gmark & \gmark & \gmark & \ymark & \rmark \\

Step-Audio-EditX \cite{yan2025step} 
& \rmark & \rmark & \rmark & \gmark & \gmark \\

\midrule

\rowcolor{myPurple!5}
\textbf{VoiceDesigner} 
& \gmark & \gmark & \gmark & \gmark & \gmark \\

\bottomrule
\end{tabular}}
\end{table}

To tackle these challenges, we introduce \textbf{VoiceDesigner}%
\footnote{Demo page: \href{https://voicedesigner-demo.github.io/}{voicedesigner-demo.github.io}.}, a unified TTV framework that supports both diverse voice generation and high-fidelity voice editing. A comparison with recent TTV systems is presented in Table~\ref{tab:ttv_model_comp}. To address data scarcity, VoiceDesigner employs two complementary simulation pipelines. A digital signal processing (DSP)-based pipeline synthesizes voices that humans cannot naturally produce (e.g., demon, ghost, monster, robot), broadening the voice distribution for both TTV generation and voice cloning. In addition, a generative simulation pipeline leverages zero-shot TTS and voice conversion models to produce diverse stylistic variations from fine-grained voice trait descriptions and paired editing data, improving generation diversity while providing supervision for voice editing. On the model side, VoiceDesigner builds upon the state-of-the-art MM-DiT~\cite{esser2024scaling, zhuo2024lumina} to unify voice generation and editing within a single framework. To better coordinate heterogeneous conditioning signals, including textual instructions, transcripts, and audio conditions, we extend MM-DiT with token-level Adaptive Layer Normalization (AdaLN) and 3D rotary positional embeddings (3D-RoPE), enabling more effective multimodal conditioning for both generation and editing.

Overall, our contributions of VoiceDesigner are threefold:

\begin{itemize}
    \item We propose a hybrid data pipeline that combines DSP techniques and generative models to support diverse human/non-human voices, rich character specifications, and editing pairs.
    
    \item We introduce an MM-DiT architecture with optimized conditioning modules to jointly model text-to-voice generation and editing, enabling high-fidelity generation and precise control. 

    \item Subjective and objective experiments demonstrate superior prompt alignments of voice description and edit instruction, while maintaining competitive perceptual quality and voice usability compared to various open-source TTV baselines.
\end{itemize}

\section{Related Work}
\label{section:related}

\subsection{Text-Guided Voice Generation}
Early works such as PromptTTS \cite{guo2023prompttts, leng2023prompttts} and PromptTTS++ \cite{shimizu2024prompttts++} introduced caption-based annotations to guide speech generation, primarily focusing on basic attributes such as gender and age. Subsequent studies explored finer-grained speaker characteristics. LibriTTS-P \cite{kawamura2024libritts} and DreamVoice \cite{hai2024dreamvoice} provide detailed descriptions of speaker timbre, while ParaSpeechCaps \cite{diwan2025scaling} offers captions covering a richer and more diverse set of speaking style attributes. SpeechCraft \cite{jin2024speechcraft} and VoxInstruct \cite{zhou2024voxinstruct} incorporate emotion classification signals to model fine-grained emotional information, though they remain limited in generation quality and subtle emotional variation. AudioBox \cite{vyas2023audiobox} combines automatically generated tags with human-annotated style labels to build scalable and diverse speech datasets, but the dataset and model details remain closed-source. More recently, CapSpeech \cite{wang2025capspeech} improves instruction-following by separating pretraining and fine-tuning, scaling pretraining with automatic annotation pipelines and releasing high-quality annotated datasets with attributes such as accent and emotion for fine-tuning. ElevenLabs further explores voice design in its voice design product \cite{elevenlabs_voice_design}, allowing users to describe desired voice characteristics via natural language. Qwen3-TTS \cite{hu2026qwen3} extends this capability by introducing character voice design within instruction-based speech generation frameworks.

\subsection{Text-Guided Voice Editing}
While substantial efforts have been made on text-instruction-guided voice generation, voice editing remains relatively underexplored. ISSE \cite{chen2025isse} proposes a voice editing task along with corresponding data simulation strategies and a model framework; however, the quality of the synthesized data and instructions remains limited, which constrains editing performance. IndexTTS-2 \cite{zhou2025indextts2} supports emotion control through emotion vector parameters within its voice cloning framework, and its official release further integrates a LLM to convert text instructions into parameter settings, enabling indirect voice style editing. More recently, Step-Audio-EditX \cite{yan2025step} introduces a dedicated framework for instruction-based voice editing that supports diverse emotions and speaking styles.

\subsection{Unified Generation and Editing Models}
Recent advances in visual generation increasingly focus on unified frameworks that support both generation and editing within a single model. While early diffusion-based approaches treated editing as conditional generation with task-specific pipelines \cite{kawar2023imagic, zhang2023adding}, more recent works \cite{xia2025dreamomni, cai2025z, wu2025qwen, xiao2025omnigen} jointly model generation targets and conditioning inputs within a shared token space. This unified representation enables consistent modeling of images and conditions such as text, masks, and reference images, supporting tasks including generation, editing, and style transfer while reducing the need for task-specific architectures.

\section{Methodology}
\label{section:method}

In this section, we first present the task formulation for text-to-voice generation and editing, covering three key aspects: voice coverage, prompt design, and editing control. We then provide a detailed description of the VoiceDesigner architecture, highlighting the design choices that enable flexible voice generation and controllable editing. Finally, we introduce the data pipeline used to construct a diverse set of voice types and their corresponding editing pairs, which serves as the foundation for training and evaluating our framework.

\subsection{Task Formulation and Voice Coverage}

\begin{figure*}[t]
    \centering
    \includegraphics[width=0.95\linewidth]{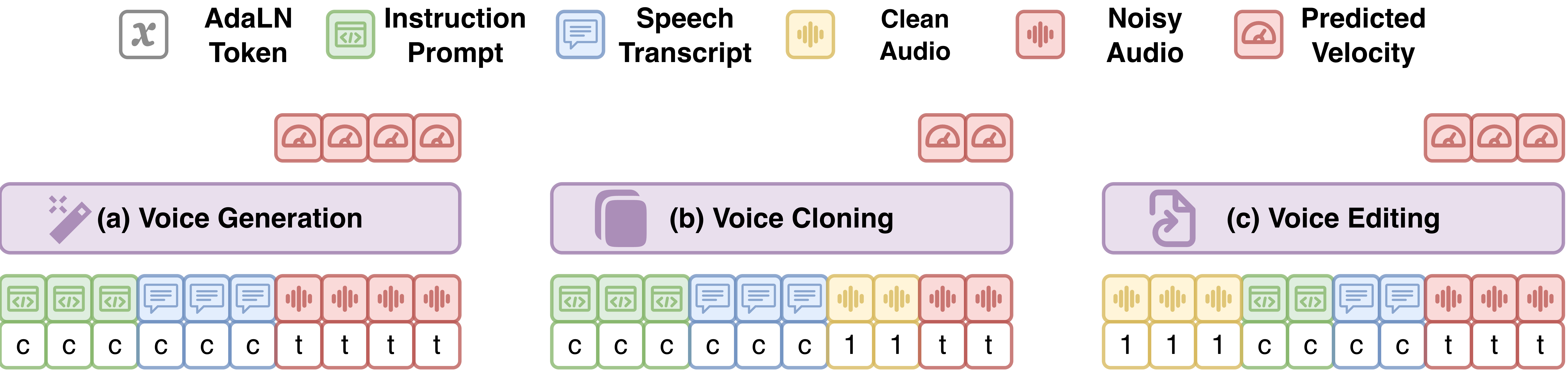}
    \caption{Generation modes of VoiceDesigner.}
    \label{fig:mode}
\end{figure*}

VoiceDesigner supports three modes: (1) voice generation from a textual voice description and transcript; (2) voice cloning from an audio reference and transcript; and (3) voice editing from an editing instruction, audio reference, and transcript. These modes are illustrated in Fig.~\ref{fig:mode}, with a detailed introduction with the model architecture in section~\ref{section:model_arch}.

For voice generation, we design a two-level voice prompt schema with examples mentioned in Section~\ref{section:introduction}. The first level specifies speaker and style traits, inspired by prior work~\cite{wang2025capspeech, hu2026qwen3}. These traits include gender, age, speaking rate, pitch, expressiveness, emotion, and accent. The second level introduces character specification, which is largely overlooked in previous studies (e.g., princess, wizard, monster, and robot), covering both human roles and non-human entities. Such character descriptors often imply distinctive prosodic patterns and are widely used in storytelling and creative production.

For voice cloning, VoiceDesigner follows the zero-shot TTS paradigm~\cite{chen2025f5}, enabling speech inpainting and continuation from a reference audio. For voice editing, VoiceDesigner treats the reference recording as an editing reference rather than a continuation prompt, enabling modification of speaker and style traits. It supports adjusting emotion, speaking style, and pitch while preserving the original speaker identity (e.g., \textit{making the speaker sound happy}), and also allows modifications to formant characteristics and special audio effects (e.g., \textit{making the voice darker}).

\subsection{Model Architecture} \label{section:model_arch}

\begin{figure*}[t]
    \centering
    \includegraphics[width=0.95\linewidth]{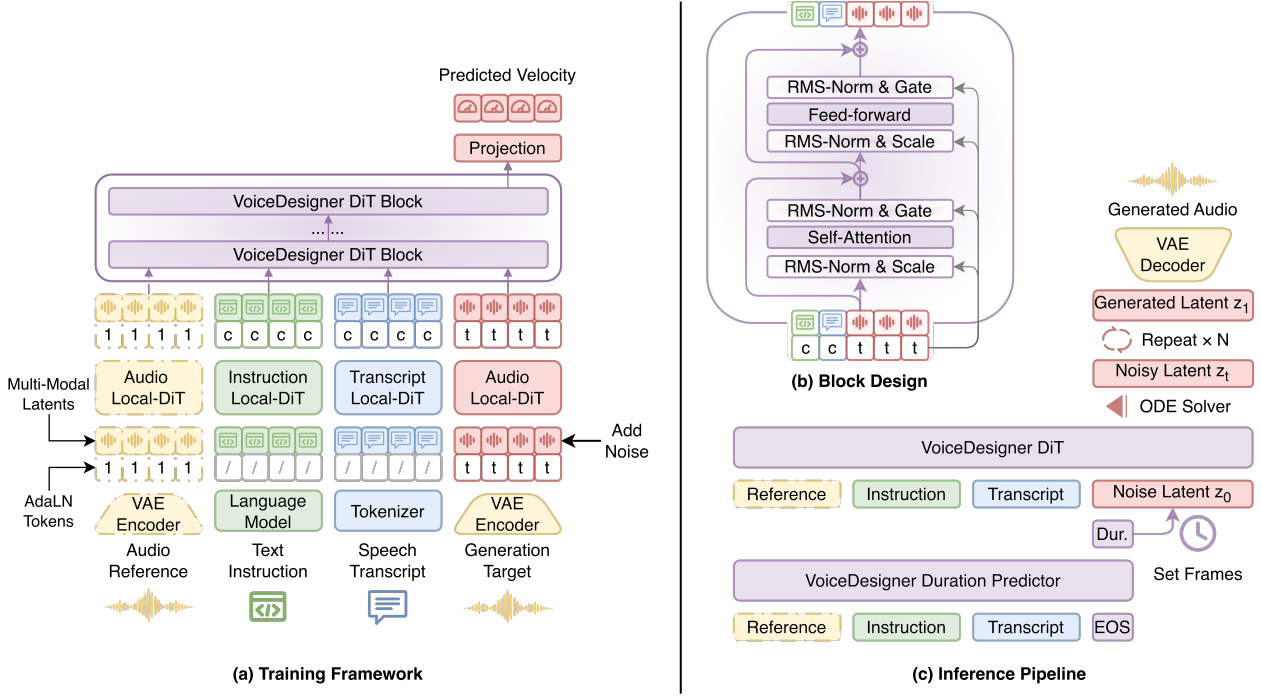}
    \caption{An overview of framework design of VoiceDesigner.}  
    \label{fig:framework}
\end{figure*}

\begin{figure*}[t]
    \centering
    \includegraphics[width=0.95\linewidth]{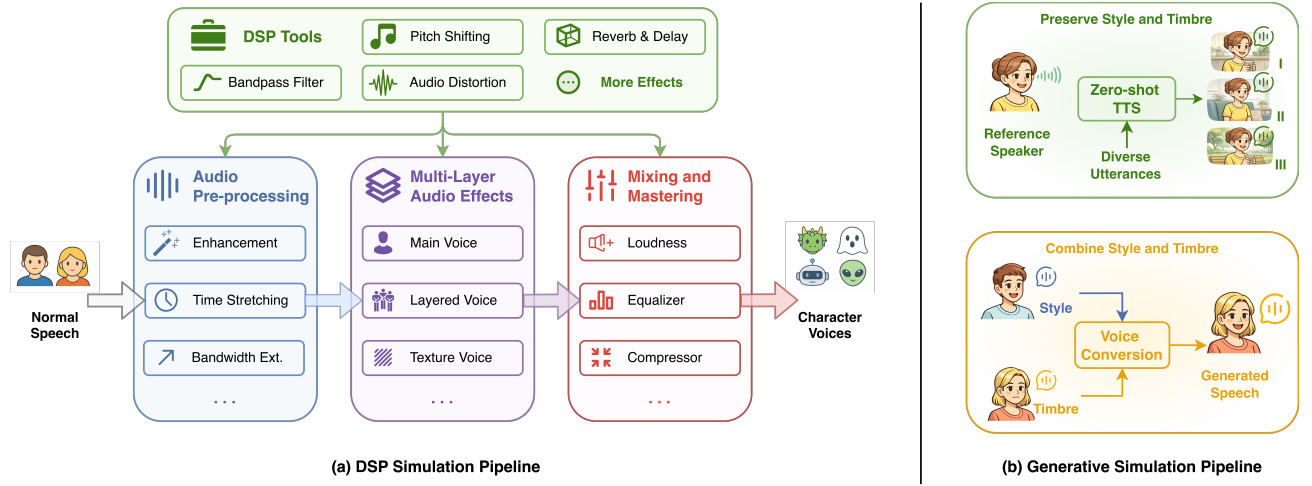}
    \caption{An overview of data simulation pipeline.}  
    \label{fig:data_sim}
\end{figure*}

\subsubsection{Overall Framework}

As shown in Fig.~\ref{fig:framework}, VoiceDesigner adopts a flow-matching diffusion model. Consistent with prior work, the model operates in the audio latent space, using DAC-VAE~\cite{kumar2023high, evans2025stable} as the encoder and decoder. The diffusion model learns to generate audio latent representations conditioned on three input modalities: transcript, instruction, and audio reference. As illustrated in Fig.~\ref{fig:mode}, different combinations of these conditioning inputs naturally define three generation modes. Voice Generation conditions on a transcript and a natural-language voice description to synthesize speech without any reference audio. Voice Cloning conditions on a transcript and a reference speech to preserve the speaker's vocal characteristics. Voice Editing conditions on a transcript, a reference speech, and an editing instruction to modify the voice attributes of the reference while preserving its linguistic content. By unifying these modes within a single diffusion model, VoiceDesigner reduces redundant training costs and enables knowledge transfer across tasks. 

As illustrated in Fig.~\ref{fig:framework} (c), during inference, we employ a sentence-level duration predictor~\cite{li2025dmospeech, yang2025simplespeech} to estimate the target speech duration. For voice generation, the predictor is conditioned on the instruction and transcript, whereas for voice editing, it is additionally conditioned on the reference audio. The predicted duration determines the length of the target latent sequence, which is initialized as Gaussian noise and subsequently denoised by the diffusion model.

\subsubsection{Diffusion Transformer}

Unlike prior diffusion-based text-to-voice systems~\cite{vyas2023audiobox, wang2025capspeech}, which primarily support one-way conditioning, VoiceDesigner handles multiple conditioning signals (instruction, transcript, audio reference) across generation, cloning, and editing modes. To unify these tasks and heterogeneous inputs, we build VoiceDesigner upon a single-stream Multimodal Diffusion Transformer (MM-DiT)~\cite{zhuo2024lumina, cai2025z} shown in Fig.~\ref{fig:framework}.

Specifically, the single-stream MM-DiT concatenates all modality tokens and processes them within a unified transformer with shared parameters. Before entering the unified transformer, tokens from each modality are first processed by a small independent transformer module. On top of this backbone, we introduce two key designs to enhance modality-aware conditioning, leading to improved performance in both voice generation and editing:

\textbf{(1) Token-level AdaLN.} In VoiceDesigner, diffusion noise is applied only to the generation target, while instruction, transcript, and audio reference remain noise-free as conditioning inputs. Unlike existing diffusion-based TTS frameworks~\cite{chen2025f5, wang2025capspeech}, where a single global diffusion timestep is sufficient because all tokens share the same noise state, VoiceDesigner jointly models multiple conditioning modalities and generation targets within a unified latent token space, resulting in heterogeneous diffusion states across token types. Consequently, a global AdaLN cannot explicitly distinguish noisy target tokens from clean conditioning tokens, forcing the model to implicitly infer their roles during denoising. This ambiguity can hinder optimization under multi-condition and multi-task training.

To address this issue, we introduce a token-level adaptive LayerNorm (AdaLN)~\cite{peebles2023scalable} inspired by diffusion forcing strategies~\cite{chen2024diffusion, song2025history}. Under the flow matching framework, each audio token is assigned a continuous time embedding $t \in [0,1]$, where $t=0$ denotes a fully noisy latent and $t=1$ a clean latent. During training, generation target tokens are assigned timesteps sampled from $[0,1]$, while audio reference tokens are fixed at $t=1$, explicitly marking them as conditioning signals. We further introduce a dedicated condition token \textbf{[C]} to distinguish instruction and transcript conditions from audio latents. This explicit per-token noise control enables the model to differentiate conditioning signals from generation targets, resulting in faster convergence and improved performance. Moreover, it naturally extends to voice cloning and speech continuation by keeping prefix speech tokens noise-free ($t=1$) while injecting diffusion noise into subsequent tokens, allowing coherent continuation conditioned on the speech prefix.

\textbf{(2) 3D-RoPE.} In TTV, three conditioning signals are involved: instruction prompts, transcripts, and audio references. When standard 1D positional embeddings are shared across all tokens~\cite{chen2025f5, yang2025simplespeech, esser2024scaling}, heterogeneous conditions are forced into a single sequential structure, which weakens modality-specific inductive bias and may hinder effective cross-modal interaction. To address this limitation, we extend rotary positional encoding (RoPE)~\cite{su2024roformer} into a three-dimensional positional space. Specifically, we define three positional dimensions, each corresponding to a different conditioning modality: instruction tokens are encoded along the $x$-axis, transcript tokens along the $y$-axis, and both clean and noisy audio tokens along the $z$-axis\footnote{A visualization of 3D-RoPE for TTV is available on the demo page.}. Aligned transcript--audio token pairs share the same $x$-coordinate to preserve their temporal correspondence, while their positions along the $y$- and $z$-axes distinguish the text and audio modalities, respectively. This 3D-RoPE design preserves both alignment-aware and modality-aware positional structures while enabling unified self-attention across all tokens, improving coordination among conditioning signals.

\subsection{Data Simulation and Augmentation}

As illustrated in Fig.~\ref{fig:data_sim}, we introduce two data simulation pipelines to construct a speech dataset of diverse voices.

\subsubsection{DSP-based Voice Simulation}

In creative media such as games and films, voices of non-human entities (e.g., monsters, ghosts, or miniature creatures) are typically created by recording human performers and applying extensive post-processing, layering, and audio effects to achieve inhuman vocal characteristics~\cite{viers2008sound}. Such voices often require acoustic transformations beyond natural human phonation, making them difficult to obtain from existing speech corpora. To enable VoiceDesigner to model these diverse non-human voice types, we develop a DSP-based audio effect synthesis pipeline. 

As shown in Fig.~\ref{fig:data_sim} (a), a standard speech recording is transformed through a sequence of audio effects, including pitch shifting, formant shifting, reverberation, equalization, band-pass filtering, and dynamic range compression. These transformations are implemented using DSP libraries such as Pedalboard~\cite{sobot_peter_2023_7817838}, librosa~\cite{mcfee2015librosa}, and Parselmouth~\cite{jadoul2018introducing}. In addition, we incorporate SiFi-GAN~\cite{yoneyama2023source} to perform pitch contour manipulation, such as flattening intonation toward monotonic speech or adjusting pitch variation ranges.

By composing these effects into structured pipelines, ordinary speech can be transformed into a wide variety of character-style voices, including dragons, demons, possessed entities, robots, astronauts, and miniature creatures. This pipeline\footnote{Representative examples, together with the complete effect chains and parameter configurations, are available on our \href{https://voicedesigner-demo.github.io/}{demo page}.} substantially expands the coverage of non-human voice types in VoiceDesigner while avoiding the need to collect dedicated recordings for each character.

\subsubsection{Generative Simulation via Voice Cloning and Conversion}

High-quality stylistic data, such as emotional, accented, and character-driven speech recordings, are often limited in both linguistic diversity and voice diversity. This restricts the generation range and generalization ability of voice models, particularly in maintaining consistent voice traits and character prosody across diverse content. To address this limitation, we propose a generative simulation pipeline that augments high-quality stylistic speech data using zero-shot text-to-speech (voice cloning) and voice conversion models, as illustrated in Fig.~\ref{fig:data_sim} (b). Voice cloning generates speech with new transcript content while preserving the reference speaker timbre and style, whereas voice conversion preserves the reference content and stylistic characteristics, such as pitch and loudness contours, while replacing the speaker timbre. 

To ensure data quality, we apply a multi-stage filtering process. (1) Linguistic accuracy: Word error rate (WER) is computed using Whisper~\cite{radford2023robust}\footnote{\href{https://huggingface.co/openai/whisper-large-v3}{https://huggingface.co/openai/whisper-large-v3}}. Samples with $\mathrm{WER}>0.1$ are discarded. For the remaining samples, the original transcript is retained if $\mathrm{WER}<0.05$; otherwise, it is replaced with the Whisper transcription. (2) Speaker and emotion consistency: Speaker and emotion embedding similarities\footnote{\href{https://github.com/BytedanceSpeech/seed-tts-eval}{https://github.com/BytedanceSpeech/seed-tts-eval}}~\cite{ma2024emotion2vec} are computed between each generated sample and the timbre reference and style reference, respectively, to evaluate speaker identity preservation and emotional consistency. Since the score distributions of these similarity metrics differ across speakers, fixed thresholds are unreliable for ranking candidates. We therefore generate two candidates for each prompt and retain the one with higher speaker similarity for voice cloning or higher emotion similarity for voice conversion. Finally, samples with speaker similarity below $0.6$ are discarded.

\section{Experimental Setup}

\subsection{Datasets}

\subsubsection{Data for Pretraining Stage} \label{sec:data_pretrain}

VoiceDesigner generates speech at a sampling rate of 48 kHz. Due to limitations in character-style data collection during the fine-tuning stage, we only focus on English for TTV generation and editing. For large-scale pretraining, we utilize four speech corpora: Emilia~\cite{he2024emilia}, Common Voice~\cite{ardila2020common}, LibriTTS-R~\cite{zen2019libritts}, and a subset of HiFiTTS-2-44.1k~\cite{langman2025hifitts}, totaling 56,165 hours. Each speech clip is annotated with speaker and vocal attributes, including age, gender, pitch, vocal expressiveness, speaking rate, and audio quality. When demographic metadata is unavailable, we employ a pre-trained age and gender estimator\footnote{\href{https://github.com/audeering/w2v2-age-gender-how-to}{https://github.com/audeering/w2v2-age-gender-how-to}}. Age is discretized into five categories: child (1–12), teen (13–19), young adult (20–39), adult (40–64), and elderly (65+).

To characterize other voice traits, we use the \textsc{PENN} library \cite{morrison2023cross} to compute speaker-level mean pitch and utterance-level pitch standard deviation, representing pitch level and tonal expressiveness. Silence and unvoiced frames are excluded for accurate estimation, and pitch values are converted to the logarithmic scale before quantization. Speaking rate is computed as the number of phonemes divided by the duration of the utterance after silence removal. Audio quality and cleanness are annotated using AudioBox-Aesthetics \cite{tjandra2025meta}. For pitch, expressiveness, and audio quality attributes, we discretize the values using percentile-based bins computed from the Emilia and HiFiTTS-2-44.1k datasets. 

Rather than generating captions directly from attribute tags, we design a set of caption templates and employ an LLM to introduce synonym variations, enabling the creation of diverse yet controlled voice descriptions.

\subsubsection{Supervised Finetuning Sets for Voice Generation} \label{sec:sft_vg}

To support diverse voice generation, we use datasets annotated with fine-grained attributes such as emotion, accent, voice timbre, and character identity. Specifically, we collect the following publicly available datasets:

\begin{itemize}
    \item \textbf{E1}: ESD English \cite{zhou2021seen}, RAVDESS \cite{livingstone2018ryerson}, and SAVEE \cite{jackson2014surrey} for recordings with common emotional categories (happiness, anger, sadness, etc.);
    \item \textbf{E2}: Expresso \cite{nguyen2023expresso}, EARS \cite{richter2024ears}, and CapSpeech-Agent \cite{wang2025capspeech} for recordings with fine-grained speaking styles (curiosity, confusion, whispering, boredom, sarcasm);
    \item VCTK and Common Voice with regional accents;
    \item DreamVoice \cite{hai2024dreamvoice} with timbre labels on LibriTTS/VCTK.
\end{itemize}

To further enhance voice diversity and support character prompts, we apply our DSP-based simulation pipeline to a bandwidth-extended version of LibriTTS-R and VCTK, synthesizing additional non-human character voice variations. In addition, we collect an internal 16-hour character voice dataset comprising 20 character identities performed by 12 professional voice actors, resulting in 360 distinct voice variations\footnote{Data descriptions and collection guidelines are provided on the \href{https://voicedesigner-demo.github.io/}{demo page} to facilitate reproducibility.}.
For all the above datasets with limited recordings per voice, we employ the generative simulation pipeline to expand scale, linguistic coverage, and emotional diversity. Specifically, we train separate voice cloning and pitch-controlled voice conversion models following designs similar to~\cite{chen2025f5, yang2025simplespeech, liu2021diffsvc, wang2025ditvc}, and use them to generate stylistically consistent speech with varied content and controlled attributes. Finally, we generate instruction-style captions using Qwen3-30B-A3B-Instruct\footnote{\href{https://huggingface.co/Qwen/Qwen3-30B-A3B-Instruct-2507}{https://huggingface.co/Qwen/Qwen3-30B-A3B-Instruct-2507}} to produce extended voice descriptions conditioned on the voice traits of each clip.

\subsubsection{Supervised Finetuning Sets for Voice Editing} 
\label{sec:sft_ve}

For instruction-guided voice editing, we construct editing pairs using both real recordings and synthetic speech generated by the generative simulation pipeline. Speaking-style editing pairs, such as emotion editing, are created from recordings of the same speaker exhibiting different styles, where either real or synthetic speech may serve as the reference or target. When real recordings are used as targets, we randomly add quality-improvement and artifact-reduction notes to the instruction prompts. We further generate editing pairs using the DSP-based simulation pipeline to simulate pitch shifting, formant shifting, and other transformations. For pitch and formant editing tasks, the processed audio is used as the reference and the original recording as the target, mitigating artifacts introduced by DSP algorithms during inference. All editing instructions are generated from curated templates with synonymous variations for each operation and speaking style, increasing linguistic diversity and improving training robustness.

\subsection{Training Setup}

\begin{table}[t]
\caption{Training configurations for different training stages.}
\label{tab:training_stages}
\centering
\begin{tabular}{lccc}
\toprule
\textbf{Training Stage}
& \textbf{Batch Size}
& \textbf{Training Steps}
& \textbf{Learning Rate} \\
\midrule
Pretraining         & $512 \times 20$ s & 400k & $1.5 \times 10^{-4}$ \\
Task Adaptation     & $384 \times 30$ s & 100k & $8 \times 10^{-5}$ \\
Quality Refinement  & $128 \times 30$ s & 10k  & $1 \times 10^{-5}$ \\
\bottomrule
\end{tabular}
\end{table}

\subsubsection{DAC-VAE and Duration Predictor Training}
We train DAC-VAE on five audio corpora: Emilia, Common Voice, VCTK, EARS, along with AudioSet~\cite{gemmeke2017audio} to further improve model generalization on general audio. All audio is resampled to 48 kHz, and the model operates with a latent frame rate of 25 Hz. The overall training setup follows the configuration used in Stable Audio \cite{evans2025stable}. The duration predictor is initialized from Qwen3-0.6B \cite{yang2025qwen3} and uses a Dasheng encoder \cite{dinkel2024scaling} to encode reference audio. A diffusion-based duration prediction head \cite{li2025dmospeech, peng2025vibevoice} is applied using the [EOS] token. The model is trained both on the pretraining and finetuning datasets.

\subsubsection{Diffusion Transformer Training}

We adopt a 1.0B-parameter\footnote{Detailed architecture parameters are available on the demo page.} MM-DiT model, where the textual instruction is encoded by T5Gemma-XL~\cite{zhang2025encoder} and the transcript is encoded by F5-TTS tokenizer~\cite{chen2025f5}. The whole training is conducted in three stages to progressively enhance generation capability and quality. The AdamW optimizer is adopted.

\textbf{Stage 1: Pretraining}. The model is trained on the speech data in Section~\ref{sec:data_pretrain} to learn voice continuation, transcript–speech alignment, and prompt–voice alignment through voice cloning and voice generation tasks.
Training is performed on 64 A100 GPUs with a maximum audio duration of 20 seconds.

\textbf{Stage 2: Task Adaptation}. The model is adapted using the datasets in section~\ref{sec:sft_vg} and~\ref{sec:sft_ve}, to further enhance the capabilities of the model in text-to-voice generation and editing. To preserve and improve speech continuation capability, we retain voice cloning training on HiFiTTS-2-44.1k together with real recordings from SFT datasets and DSP-simulated character voices. Training is conducted on 64 A100 GPUs with a maximum audio duration of 30 seconds.

\textbf{Stage 3: Quality Refinement}. In the final stage, the model is further fine-tuned for 10k steps using only real recordings and a small set of high-quality DSP-simulated data. Training is conducted on 16 A100 GPUs with a maximum audio duration of 30 seconds.

\section{Results and Analysis}
This section presents the experimental setup and evaluation results. We first evaluate VoiceDesigner on text-to-voice generation for both natural and character voices. We then assess its voice reuse and editing capabilities through zero-shot voice cloning and voice editing. Finally, we present ablation studies and qualitative analyses to examine the contributions of individual components.
\label{section:exps}

\subsection{Evaluation on Voice Generation}
\label{sec:eval_ttv}

\begin{table*}[t]
\caption{Voice generation performance across objective evaluations and two subjective listening tests.
$^*$ indicates the commercial closed-source systems accessed via API.
All MOS results are reported with 95\% confidence intervals.}
\label{tab:all}
\centering
\small
\setlength{\tabcolsep}{5pt}

\resizebox{\textwidth}{!}{
\begin{tabular}{lcccccccc}
\toprule
\multirow{2}{*}{\textbf{Model}} 
& \multicolumn{2}{c}{\textbf{Objective Evaluation}}
& \multicolumn{3}{c}{\textbf{Subjective: TTV-Traits}}
& \multicolumn{3}{c}{\textbf{Subjective: TTV-Character}} \\
\cmidrule(lr){2-3} \cmidrule(lr){4-6} \cmidrule(lr){7-9}

& \textbf{WER (\%)}$\downarrow$ 
& \textbf{Style-ACC (\%)}$\uparrow$
& \textbf{MOS-C}$\uparrow$ 
& \textbf{MOS-U}$\uparrow$ 
& \textbf{MOS-Q}$\uparrow$
& \textbf{MOS-C}$\uparrow$ 
& \textbf{MOS-U}$\uparrow$ 
& \textbf{MOS-Q}$\uparrow$ \\

\midrule
ElevenLabs-TTV API$^*$ \cite{elevenlabs_voice_design} 
& 1.68 
& \underline{0.58}
& $\mathbf{4.00} \pm 0.06$
& $\mathbf{4.10} \pm 0.05$
& $3.97 \pm 0.05$
& $\mathbf{4.26} \pm 0.05$
& $\mathbf{4.24} \pm 0.05$
& $\mathbf{4.23} \pm 0.05$ \\

CapSpeech-NAR \cite{wang2025capspeech}
& 2.12
& 0.57
& $3.26 \pm 0.07$
& $3.34 \pm 0.06$
& $3.96 \pm 0.05$
& -- & -- & -- \\

Qwen3TTS-VoiceDesign \cite{hu2026qwen3}
& \underline{1.39}
& 0.50
& $3.10 \pm 0.08$
& $3.57 \pm 0.06$
& $\mathbf{4.20} \pm 0.05$
& $3.59 \pm 0.07$
& $\underline{4.03} \pm 0.05$
& $\underline{4.19} \pm 0.05$ \\

\midrule
\rowcolor{myPurple!5}
\textbf{VoiceDesigner}
& \textbf{1.22}
& \textbf{0.66}
& $3.93 \pm 0.06$
& $3.57 \pm 0.06$
& $3.96 \pm 0.05$
& $3.71 \pm 0.06$
& $3.72 \pm 0.06$
& $3.60 \pm 0.06$ \\

\rowcolor{myPurple!5}
\textbf{VoiceDesigner (+Enhance)}
& -- & --
& $\underline{3.94} \pm 0.06$
& $\underline{3.64} \pm 0.06$
& $\underline{4.17} \pm 0.05$
& $\underline{3.76} \pm 0.06$
& $3.93 \pm 0.05$
& $3.92 \pm 0.05$ \\

\bottomrule
\end{tabular}
}
\end{table*}

We conduct both objective and subjective evaluations\footnote{All participants gave informed consent before the listening tests.} on TTV generation performance. For objective evaluation, For objective evaluation, we report \textbf{WER} computed using Whisper~\cite{radford2023robust} to measure the linguistic accuracy of the generated speech. We further follow the CapSpeech evaluation tool\footnote{\href{https://github.com/WangHelin1997/CapSpeech}{https://github.com/WangHelin1997/CapSpeech}} to adopt a classification-based framework to measure style accuracy (\textbf{Style-ACC}) of the model generation. In this framework, pretrained classifiers are used to extract speech attributes from generated audio, including gender, age, pitch, emotion~\cite{ma2024emotion2vec}, and accent~\cite{accent-classifier-6class}. The predicted attributes are compared with the prompt descriptions to evaluate prompt–voice alignment. For pitch and other prosodic features, such as tone variation and speaking rate, we retain three discrete levels (low, mid, high) based on the distributions observed in HiFiTTS-2 and Emilia. To support this evaluation, we construct a benchmark dataset containing 520 prompt–voice pairs, where accent and emotion are included as they can be reliably evaluated using existing classification models. The specifications of these prompts are: (1) 160 prompts of only fundamental speaker attributes of gender, age, pitch, expression, and speaking rate; (2) 120 prompts that additionally include accent; (3) 120 prompts that combine basic speaker attributes with emotions supported by emotion classification models; and (4) 120 prompts with all above attributes. All prompts are first generated by GPT 5.1 then verified by humans. 

For subjective evaluation, we design two listening test benchmarks, with samples available in the demo page:

\begin{itemize}
    \item TTV-Traits: this benchmark is settled on human voices, with voice prompts of different speaker and style attributes, to evaluate the performance of generating human voices with diverse styles and timbres. The benchmark comprises of 75 prompt-voice pairs evenly from 15 emotions or styles. The age, gender, accent and pitch are randomly assigned. 
    \item TTV-Character: this benchmark includes both human and non-human voices, with text prompts specifying character traits and voice attributes (e.g., emotion). It comprises 150 prompt--voice samples covering 50 characters.
\end{itemize}

We adopt three Mean Opinion Score (MOS) metrics, each rated on a 5-point scale: (1) \textbf{MOS-C}, which measures consistency with the instruction prompt; (2) \textbf{MOS-U}, which evaluates the production usability of the generated voice in terms of character portrayal and vocal prosody; and (3) \textbf{MOS-Q}, which assesses perceptual audio quality. For both experiments, 100 trained listeners participated in the listening tests, with each listener evaluating 25 recordings. Every recording was rated by at least four different listeners. Unless otherwise specified, all subsequent subjective evaluations follow the same 5-point MOS protocol and use the same pool of trained listeners.

Table~\ref{tab:all} summarizes the results of the three TTV generation evaluations across four models. We compare VoiceDesigner with two recent open-source systems (CapSpeech-NAR~\cite{wang2025capspeech} and Qwen3TTS-VoiceDesign~\cite{hu2026qwen3}) and one leading commercial model (ElevenLabs-TTV API\footnote{\href{https://elevenlabs.io/}{https://elevenlabs.io/}}). Based on our observations, the raw outputs of VoiceDesigner occasionally preserve environmental ambience and low-level background noise, whereas ElevenLabs and Qwen3TTS appear to employ proprietary post-processing (e.g., denoising and dereverberation) to improve perceptual quality. For a fair comparison, we additionally report a post-processed variant of VoiceDesigner using an enhancement tool\footnote{\href{https://podcast.adobe.com/en/enhance}{https://podcast.adobe.com/en/enhance}}. Editing and voice cloning are based on input audio and thus require no additional enhancement. This provides an estimate of the perceptual quality attainable with external enhancement while keeping the underlying generation model unchanged. 

From the objective evaluation results, VoiceDesigner achieves the strongest prompt–style alignment (Style-ACC) among all systems. ElevenLabs-TTV API and CapSpeech-NAR show competitive performance while Qwen3TTS lags behind. Regarding intelligibility, VoiceDesigner achieves the lowest word error rate. These objective results demonstrate that VoiceDesigner provides strong style alignment while maintaining competitive transcription accuracy.

Across the two subjective evaluations, VoiceDesigner demonstrates strong prompt–voice alignment (MOS-C), outperforming all open-source baselines on both benchmarks, while remaining below the leading commercial system. In terms of voice usability (MOS-U), VoiceDesigner is competitive with Qwen3TTS-VoiceDesign and surpasses CapSpeech-NAR, exhibiting stable prosody, appropriate pausing, and expressive delivery across both human and non-human voices. Regarding perceptual quality (MOS-Q), Qwen3TTS-VoiceDesign produces particularly clean outputs. However, with the enhancement module applied, VoiceDesigner significantly improves audio quality, narrowing the gap while maintaining its advantage in instruction alignment.

Overall, VoiceDesigner achieves superior prompt–voice coherence among open-source systems, while maintaining competitive usability and perceptual quality. Although a performance gap remains compared to ElevenLabs, the MOS-C score shows encouraging proximity, suggesting that following the same data simulation and modeling paradigms could help close this gap with additional real-world training data.  

\subsection{Evaluation on Voice Cloning}
\label{sec:eval_tts}

\begin{table}[t]
\caption{Voice cloning performance on the Seed-TTS test-en benchmark, where $^{*}$ denotes closed-source models; results are from official technical reports.}
\label{tab:zero_shot_tts}
\centering
\small
\setlength{\tabcolsep}{13pt}
\resizebox{\linewidth}{!}{
\begin{tabular}{lcc}
\toprule
\textbf{Model} & \textbf{WER (\%) $\downarrow$} & \textbf{SIM-o $\uparrow$} \\
\midrule
Ground Truth & 2.06 & 0.734 \\
\midrule
SeedTTS$^{*}_{\text{DiT}}$ \cite{anastassiou2024seed} & \underline{1.73} & \textbf{0.790} \\
E2-TTS \cite{chen2025f5, eskimez2024e2} & 2.19 & 0.710 \\
F5-TTS \cite{chen2025f5} & 1.83 & 0.670 \\
CosyVoice-3-1.5B$^{*}$ \cite{du2025cosyvoice} & 2.21 & 0.720 \\
CosyVoice-3-0.5B \cite{du2025cosyvoice} & 2.02 & 0.718 \\
IndexTTS-2 \cite{zhou2025indextts2} & 2.23 & 0.706 \\
\midrule
\rowcolor{myPurple!5}
\textbf{VoiceDesigner} & \textbf{1.70} & \underline{0.757} \\
\bottomrule
\end{tabular}}
\end{table}

\begin{table}[t]
\caption{Subjective evaluation for challenging voice cloning.}
\label{tab:challenge_vc}
\centering
\small
\setlength{\tabcolsep}{8pt}
\resizebox{\linewidth}{!}{
\begin{tabular}{lcc}
\toprule
\textbf{Model} 
& \textbf{MOS-T $\uparrow$} 
& \textbf{MOS-S $\uparrow$} \\
\midrule
CosyVoice-3-0.5B \cite{du2025cosyvoice}
& $2.48 \pm 0.08$ 
& $2.63 \pm 0.08$ \\

IndexTTS-2 \cite{zhou2025indextts2}
& $\underline{3.85} \pm 0.07$ 
& $\underline{3.94} \pm 0.06$ \\
\midrule
\rowcolor{myPurple!5}
\textbf{VoiceDesigner} 
& $\mathbf{3.93} \pm 0.07$ 
& $\mathbf{4.00} \pm 0.06$ \\
\bottomrule
\end{tabular}}
\end{table}

To evaluate voice cloning capability, we follow the commonly used zero-shot TTS evaluation protocol on the Seed-TTS test-en benchmark \cite{anastassiou2024seed} for human voices and compare our model with recent state-of-the-art systems. We use speaker similarity (\textbf{SIM-o}) between generated and original recording to assess speaker preservation, and \textbf{WER} to evaluate intelligibility and transcription accuracy.

Meanwhile, to evaluate the model's ability to clone special character voices, we construct a dedicated test set consisting of 20 voice--utterance pairs covering a diverse set of challenging character voices, including dragons, monsters, robots, and other non-human characters. We compare our model with CosyVoice-3 and IndexTTS-2, which demonstrate strong performance on voice cloning benchmarks across character-based human and non-human voices, while other baselines are dropped due to their inability to clone them. Similarly, we conduct a subjective listening test using a 5-point MOS scale to assess: (1) \textbf{MOS-T}, which measures voice timbre similarity between the generated speech and the reference recording; and (2) \textbf{MOS-S}, which measures speaking-style similarity between the generated speech and the reference recording. A total of 110 trained listeners each evaluated 25 recordings, with every recording rated by at least four listeners.

As shown in Table~\ref{tab:zero_shot_tts}, VoiceDesigner achieves the lowest WER among all compared systems, indicating high speech generation accuracy. Its speaker similarity score also surpasses most recent zero-shot TTS models and remains close to that of the closed-source SeedTTS, demonstrating strong voice cloning performance on natural human voices.

Table~\ref{tab:challenge_vc} further shows that, based on subjective evaluations, VoiceDesigner achieves the best performance in both voice timbre and speaking-style similarity. It surpasses both CosyVoice-3 and IndexTTS-2, further demonstrating its strong capability in cloning special character voices.

\subsection{Evaluation on Voice Editing}

\begin{table}[t]
\caption{Evaluation results for voice editing performance.}
\label{tab:voice_editing}
\centering
\small
\setlength{\tabcolsep}{13pt}
\begin{tabular}{lcc}
\toprule
\textbf{Model} 
& \textbf{MOS-E $\uparrow$} 
& \textbf{SIM-t $\uparrow$} \\
\midrule
Ground Truth & $4.147 \pm 0.056$ & -- \\
\midrule
IndexTTS-2 \cite{zhou2025indextts2} & $\underline{3.494} \pm 0.072$ & \underline{0.856} \\
Step-Audio-EditX \cite{yan2025step} & $3.333 \pm 0.064$ & 0.734 \\
\midrule
\rowcolor{myPurple!5}
\textbf{VoiceDesigner} & $\textbf{4.129} \pm 0.054$ & \textbf{0.884} \\
\bottomrule
\end{tabular}
\end{table}

\begin{figure}[t]
    \centering
    \includegraphics[width=0.925\linewidth]{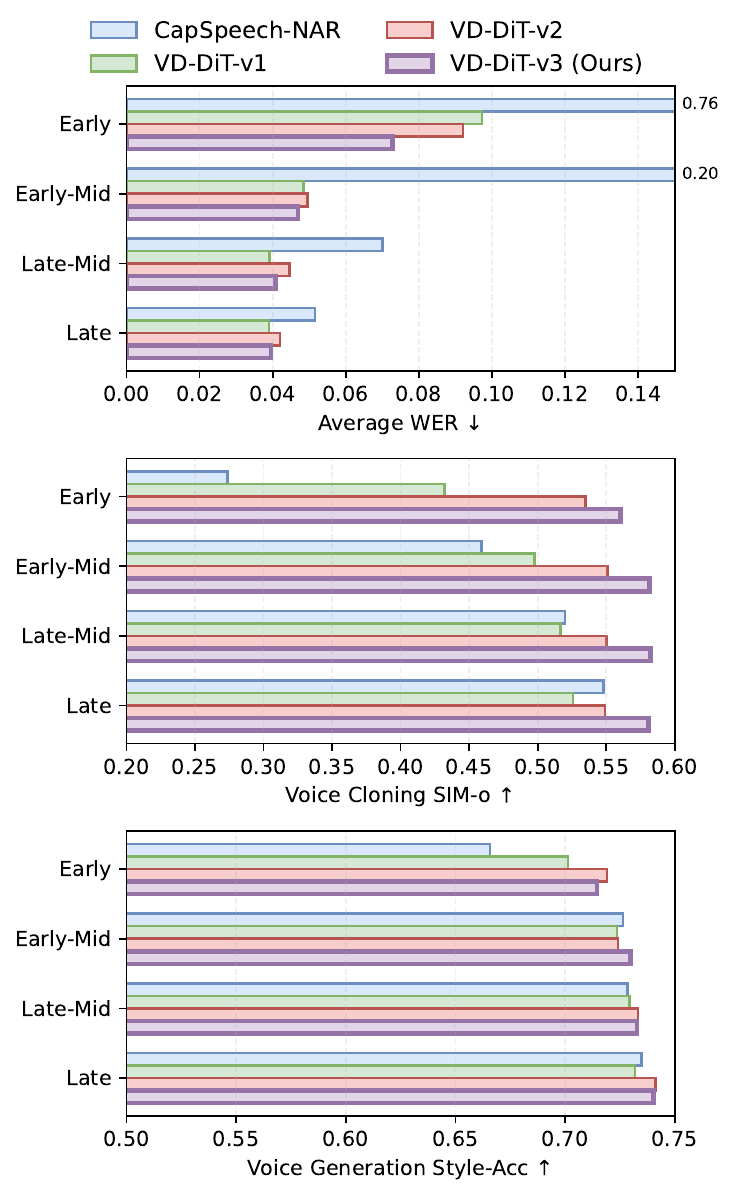}
    \caption{Ablation study on the model architecture.}    
    \label{fig:architecture_ablation}
\end{figure}

\begin{figure}[t]
    \centering
    \includegraphics[width=0.925\linewidth]{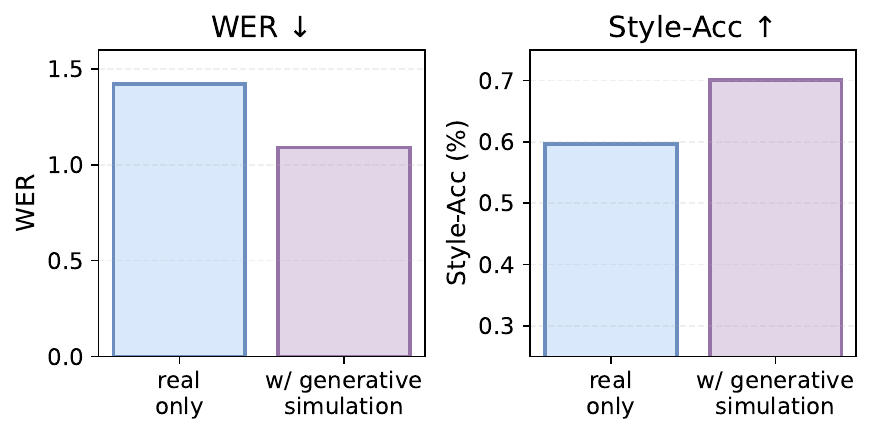}
    \caption{Ablation study on generative data augmentation.}
\label{fig:ablation-data}
\end{figure}

To evaluate voice editing performance, we conduct subjective experiments focused on emotion and speaking style editing, as most existing voice editing approaches primarily target such modifications\footnote{For editing capabilities not supported by baselines (e.g., pitch modification), see additional examples on our demo page.}. The evaluation is based on an internal expressive speech dataset of professional voice actors. Each speaker provides utterances covering a wide range of speaking styles and emotional expressions. This dataset is never trained in VoiceDesigner across all three stages.

Following the evaluation protocol of Step-Audio-EditX, we randomly construct 100 voice editing pairs spanning emotional and stylistic transformation scenarios while excluding ambiguous or weakly defined styles. These pairs are derived from 50 distinct source utterances spoken by 10 speakers, with each source recording edited into two different target styles. We compare VoiceDesigner with Step-Audio-EditX and the text-prompt emotion control mode of IndexTTS-2.

For objective evaluation, we measure voice similarity (\textbf{SIM-t}) between the edited speech and the ground-truth recording of the target style using WeSpeaker ResNet-34 embedding similarity~\cite{wang2023wespeaker}, which has been shown to be more robust and effective than traditional speaker embeddings such as ECAPA~\cite{desplanques2020ecapa} and X-vector~\cite{snyder2018xvectors}. We further conduct a subjective listening test using a 5-point MOS scale to assess \textbf{MOS-E} for editing instruction following, which measures how accurately the edited speech reflects the specified modification relative to the original recording. The evaluation involves 110 trained listeners, each rating 25 recordings, with every recording receiving at least four ratings.

As shown in Table~\ref{tab:voice_editing},
VoiceDesigner surpasses both Step-Audio-EditX and IndexTTS-2 in voice similarity and instruction-following accuracy, demonstrating strong capability in fine-grained voice style editing. In contrast, Step-Audio-EditX occasionally introduces audible artifacts, while both baselines exhibit inconsistent performance when editing certain voice styles.

\subsection{Ablation Studies and Additional Analysis}

\subsubsection{Architecture Comparison}

For diffusion transformer architecture comparison, we evaluate several design variants, including CapSpeech-NAR, which extends F5-TTS by introducing cross-attention to support TTV modeling, together with our proposed architectures:
\begin{itemize}
    \item VD-DiT-v1: original single-stream MM-DiT~\cite{zhuo2024lumina, cai2025z};
    \item VD-DiT-v2: MM-DiT with token-level AdaLN;
    \item VD-DiT-v3: VD-DiT-v2 with 3D-RoPE.
\end{itemize}

Under the voice cloning setting, for CapSpeech-NAR and VD-DiT-v1, we follow the speaker conditioning strategy adopted in F5-TTS \cite{chen2025f5} and E2-TTS \cite{eskimez2024e2}, where speaker embeddings are concatenated with the noisy latent along the channel dimension and masked regions are filled with empty tokens. For VD-DiT-v1 and VD-DiT-v2, speaker condition tokens are assigned a flow-matching time step of 1, such that Gaussian noise is not injected into these tokens, while diffusion noise is applied to all remaining generation tokens.

We conduct ablation studies using a 5,000-hour subset of the HiFiTTS-2 dataset. 200 speakers are excluded from training and reserved for evaluation. Model performance is evaluated under both zero-shot TTS (voice cloning) and TTV scenarios. For the TTV setting, prompts are generated following the Stage-1 training strategy, consistent with the protocols adopted in CapSpeech. For voice cloning evaluation, we construct 400 reference–utterance pairs from the excluded speakers. And we construct another 400 prompt–utterance pairs to evaluate TTV performance. Following the evaluation protocol described in Sections~\ref{sec:eval_ttv} and~\ref{sec:eval_tts}, we report \textbf{WER} and \textbf{SIM-o} for the voice cloning setting, and \textbf{Style-Acc} to measure the consistency of fundamental speaker traits under instruction-guided generation in the TTV scenario. For fair comparison, we adjust the model width and layers to ensure the approximately 0.3B parameters. All models are trained using 16 A100 GPUs with a global batch size of 128 and a learning rate of $1\times10^{-4}$.

As shown in Figure~\ref{fig:architecture_ablation}, we visualize the metric values of different training steps to evaluate the convergence efficiency, namely Early (50K steps), Early-Mid (100K), Late-Mid (150K), and Late (200K). VD-DiT-v1 converges faster and achieves significantly lower WER than CapSpeech-NAR, likely because CapSpeech shares a unified positional space for audio and text tokens and therefore requires more training to disentangle and align the two modalities. In voice cloning task particularly for speaker similarity and voice generation, VD-DiT-v1 performs better at the early stage but becomes slightly weaker than CapSpeech-NAR later in training. With token-level AdaLN, VD-DiT-v2 explicitly distinguishes conditional tokens, leading to substantially improved convergence speed and final performance on speaker similarity and voice generation, with only a minor increase in WER. By introducing 3D-RoPE to assign distinct positional patterns to instruction, speech content, and audio target tokens, VD-DiT-v3 reduces interference among heterogeneous tokens and improves structural alignment across sequences, resulting in significant gains in speaker similarity, further reduction in WER compared to VD-DiT-v2, and comparable performance in voice generation.

\subsubsection{Generative Simulation Data Augmentation}
To verify the effectiveness of the generative simulation pipeline via voice cloning and voice conversion for data augmentation, we conduct experiments using the public emotional speech corpora \textbf{E1} and \textbf{E2} described in section~\ref{sec:data_pretrain}. Although these datasets provide human-annotated emotion labels, they are limited in terms of total duration and linguistic diversity. In the ablation study, we follow the same augmentation procedure described in Section~\ref{sec:sft_vg}, which employs voice cloning and voice conversion to generate additional training data. 

We initialize the model from a Stage-1 checkpoint and further fine-tune it on the voice generation task. For evaluation, we adopt a simplified version of the objective evaluation dataset introduced in section~4.2.1, excluding accent-related attributes. Model performance is compared using \textbf{Style-Acc} and \textbf{WER}. Specifically, we evaluate models trained with (1) real data only; and (2) real data augmented with generative samples. Figure~\ref{fig:ablation-data} shows that generative data augmentation consistently improves performance in voice generation, with better linguistic fidelity and instruction adherence. These results demonstrate the effectiveness of generative augmentation in improving both content accuracy and style controllability.

\subsubsection{Runtime Analysis} 
Runtime performance is evaluated using the Real-Time Factor (RTF) on a single NVIDIA RTX 4090 GPU. The RTF is averaged over 200 test utterances with an average duration of approximately 15\,s. VoiceDesigner achieves RTFs of 0.36 for voice generation, comparable to CapSpeech (0.33) and substantially faster than Qwen3-TTS (1.21), and 0.42 for voice editing, outperforming Step-Audio-EditX (0.80) and IndexTTS2 (0.72).

\section{Conclusion}

We presented VoiceDesigner, a unified framework for text-to-voice generation and editing that supports diverse voice types and flexible editing instructions. To address the limited voice diversity and weak editing capability of existing TTV systems, we proposed a hybrid data construction pipeline to expand coverage of both human and non-human voices. We further introduced a MM-DiT-based architecture with modality-aware conditioning designs to jointly model voice generation, cloning, and instruction-guided editing within a single framework. Future work will focus on expanding real-world character-style datasets, improving high-fidelity modeling, and further narrowing the gap with commercial systems through enhanced data diversity and model scaling.

\newpage
 
\bibliographystyle{IEEEtran}
\bibliography{mybib}

\vfill

\end{document}